\documentclass{SciPost}

\usepackage[utf8]{inputenc}
\usepackage[T1]{fontenc}
\usepackage{amsmath,latexsym,graphicx}
\usepackage[bitstream-charter]{mathdesign}

\fancypagestyle{SPstyle}{
\fancyhf{}
\lhead{\colorbox{scipostblue}{\bf \color{white} ~SciPost Physics }}
\rhead{{\bf \color{scipostdeepblue} ~Submission }}

\fancyfoot[C]{\textbf{\thepage}}
}

\DeclareSymbolFont{usualmathcal}{OMS}{cmsy}{m}{n}
\DeclareSymbolFontAlphabet{\mathcal}{usualmathcal}

\hypersetup{
    colorlinks,
    linkcolor={red!50!black},
    citecolor={blue!50!black},
    urlcolor={blue!80!black}
}

\begin{document}

\pagestyle{SPstyle}

\begin{center}{\Large \textbf{\color{scipostdeepblue}{
  On the topology of solutions to random continuous constraint satisfaction problems\\
}}}\end{center}

\begin{center}
  \textbf{Jaron Kent-Dobias}
\end{center}

\begin{center}
Istituto Nazionale di Fisica Nucleare, Sezione di Roma I, Italy
\\
ICTP South American Institute for Fundamental Research, São Paulo, Brazil
\\
Instituto de Física Teórica, Universidade Estadual Paulista ``Júlio de Mesquita Filho'', São Paulo, Brazil
\\[\baselineskip]
\href{mailto:jaron@ictp-saifr.org}{\small jaron@ictp-saifr.org}
\end{center}

\section*{\color{scipostdeepblue}{Abstract}}
\textbf{\boldmath{%
We consider the set of solutions to $M$ random polynomial equations whose $N$
variables are restricted to the $(N-1)$-sphere. Each equation has independent
Gaussian coefficients and a target value $V_0$. When solutions exist, they form
a manifold. We compute the average Euler characteristic of this manifold in the
limit of large $N$, and find different behavior depending on the target value
$V_0$, the ratio $\alpha=M/N$, and the variances of the coefficients. We divide
this behavior into five phases with different implications for the topology of
the solution manifold. When $M=1$ there is a correspondence between this
problem and level sets of the energy in the spherical spin glasses. We
conjecture that the transition energy dividing two of the topological phases
corresponds to the energy asymptotically reached by gradient descent from a
random initial condition, possibly resolving an open problem in
out-of-equilibrium dynamics. However, the quality of the available data leaves
the question open for now.
}
}

\vspace{\baselineskip}

\noindent\textcolor{white!90!black}{%
\fbox{\parbox{0.975\linewidth}{%
\textcolor{white!40!black}{\begin{tabular}{lr}%
  \begin{minipage}{0.6\textwidth}%
    {\small Copyright attribution to authors. \newline
    This work is a submission to SciPost Physics. \newline
    License information to appear upon publication. \newline
    Publication information to appear upon publication.}
  \end{minipage} & \begin{minipage}{0.4\textwidth}
    {\small Received Date \newline Accepted Date \newline Published Date}%
  \end{minipage}
\end{tabular}}
}}
}


\vspace{10pt}
\noindent\rule{\textwidth}{1pt}
\tableofcontents
\noindent\rule{\textwidth}{1pt}
\vspace{10pt}

\section{Introduction}

Constraint satisfaction problems seek configurations that simultaneously
satisfy a set of equations, and form a basis for thinking about problems as
diverse as neural networks \cite{Mezard_2009_Constraint}, granular materials
\cite{Franz_2017_Universality}, ecosystems \cite{Altieri_2019_Constraint}, and
confluent tissues \cite{Urbani_2023_A}. All but the last of these examples deal
with sets of inequalities, while the last considers a set of equality
constraints. Inequality constraints are familiar in situations like zero-cost
solutions in neural networks with ReLu activations and stable equilibrium in the
forces between physical objects. Equality constraints naturally appear in the
zero-gradient solutions to overparameterized smooth neural networks and in vertex models of tissues.

In problems ranging from toy models \cite{Baldassi_2016_Unreasonable, Baldassi_2019_Properties} to real deep neural networks \cite{Goodfellow_2014_Qualitatively, Draxler_2018_Essentially, Frankle_2020_Revisiting, Vlaar_2022_What, Wang_2023_Plateau}, there is great interest in characterizing structure in the
set of solutions, which can influence the behavior of algorithms trying
to find them \cite{Beneventano_2023_On}. Here, we show how topological information about
the set of solutions can be calculated in a simple problem of satisfying random
nonlinear equalities. This allows us to reason about the connectivity and structure of the
solution set. The topological properties revealed by this calculation yield
surprising results for the well-studied spherical spin glasses, where a
topological transition thought to occur at a threshold energy $E_\text{th}$
where marginal minima are dominant is shown to occur at a different energy
$E_\text{sh}$. We conjecture that this difference resolves an outstanding
problem with the out-of-equilibrium dynamics in these systems.

We consider the problem of finding configurations $\mathbf x\in\mathbb R^N$
lying on the $(N-1)$-sphere $\|\mathbf x\|^2=N$ that simultaneously satisfy $M$
nonlinear constraints $V_k(\mathbf x)=V_0$ for $1\leq k\leq M$ and some
constant $V_0\in\mathbb R$. The nonlinear constraints are taken to be centered
Gaussian random functions with covariance
\begin{equation} \label{eq:covariance}
  \overline{V_i(\mathbf x)V_j(\mathbf x')}
  =\delta_{ij}f\left(\frac{\mathbf x\cdot\mathbf x'}N\right)
\end{equation}
for some choice of function $f$. When the covariance function $f$ is polynomial, the
$V_k$ are also polynomial, with a term of degree $p$ in $f$ corresponding to
all possible terms of degree $p$ in the $V_k$. One can explicitly construct functions that satisfy \eqref{eq:covariance} by taking
\begin{equation}
  V_k(\mathbf x)
  =\sum_{p=0}^\infty\frac1{p!}\sqrt{\frac{f^{(p)}(0)}{N^p}}
  \sum_{i_1\cdots i_p}^NJ^{(k,p)}_{i_1\cdots i_p}x_{i_1}\cdots x_{i_p}
\end{equation}
where the elements of the tensors $J^{(k,p)}$ are independently distributed
unit normal random variables. The
series coefficients of $f$ therefore control the variances of the random coefficients
in the polynomials $V_k$. When $M=1$, this problem corresponds to
finding the level set of a spherical spin glass at energy density
$E=V_0/\sqrt{N}$.

This problem or small variations thereof have attracted attention recently for
their resemblance to encryption, least-squares optimization, and vertex models
of confluent tissues \cite{Fyodorov_2019_A, Fyodorov_2020_Counting,
  Fyodorov_2022_Optimization, Tublin_2022_A, Vivo_2024_Random, Urbani_2023_A,
  Kamali_2023_Dynamical, Kamali_2023_Stochastic, Urbani_2024_Statistical,
Montanari_2023_Solving, Montanari_2024_On, Kent-Dobias_2024_Conditioning,
Kent-Dobias_2024_Algorithm-independent}. In each of these cases, the authors
studied properties of the cost function
\begin{equation} \label{eq:cost}
  \mathscr C(\mathbf x)=\frac12\sum_{k=1}^M\big[V_k(\mathbf x)-V_0\big]^2
\end{equation}
which achieves zero only for configurations that satisfy all the constraints.
Introduced in Ref.~\cite{Fyodorov_2019_A}, the existence of solutions and the
geometric structure of the cost function were studied for the problem with linear $V_k$ in
a series of papers \cite{Fyodorov_2019_A, Fyodorov_2020_Counting,
Fyodorov_2022_Optimization} and later reviewed \cite{Vivo_2024_Random}. Some
work on the equilibrium measure of the cost function with nonlinear $V_k$
was made in Ref.~\cite{Tublin_2022_A}, and the problem was solved in
Ref.~\cite{Urbani_2023_A}. Subsequent work has studied varied dynamics applied
to the cost function, including gradient descent, Hessian descent, Langevin,
stochastic gradient descent, and approximate message passing
\cite{Kamali_2023_Dynamical, Kamali_2023_Stochastic, Montanari_2023_Solving,
Montanari_2024_On}. Finally, some progress has been made on aspects of the
geometric structure of the cost function with nonlinear $V_k$
\cite{Kent-Dobias_2024_Conditioning, Kent-Dobias_2024_Algorithm-independent}.

From the perspective of the cost function, the set of solutions looks like a network of flat canyons at the bottom of the cost landscape.
Here we dispense with the cost function and study the set of solutions
directly. This set can be written as
\begin{equation}
  \Omega=\big\{\mathbf x\in\mathbb R^N\mid \|\mathbf x\|^2=N,V_k(\mathbf x)=V_0
  \;\forall\;k=1,\ldots,M\big\}
\end{equation}
Because the constraints are all smooth functions, $\Omega$ is almost always a manifold without singular points.\footnote{The conditions for a singular point are that
  $0=\frac\partial{\partial\mathbf x}V_k(\mathbf x)$ for all $k$. This is
  equivalent to asking that the constraints $V_k$ all have a stationary point at
  the same place. When the $V_k$ are independent and random, this is vanishingly
  unlikely, requiring $NM+1$ independent equations to be simultaneously satisfied.
  This means that different connected components of the set of solutions do not
intersect, nor are there self-intersections, without extraordinary fine-tuning.}
We study the topology of the manifold $\Omega$ by computing its
average Euler characteristic, a topological invariant whose value puts
constraints on the manifold's structure. The topological phases determined
by this measurement are distinguished by the size and sign of the Euler
characteristic, and the distribution in space of its constituents.

In Section~\ref{sec:euler.1} we describe how to calculate the average Euler
characteristic, how to interpret the results of that calculation, and what
topological phases are implied. In Section~\ref{sec:ssg} we examine some
implications of these results for dynamic thresholds in the spherical spin glasses.
Finally, in Section~\ref{sec:conclusion} we make some concluding remarks. Many
of the details of the calculations in the middle sections are found in
Appendices~A--D.

\section{The average Euler characteristic}
\label{sec:euler.1}

\subsection{Definition and derivation}

The Euler characteristic $\chi$ of a manifold is a topological invariant \cite{Hatcher_2002_Algebraic}. It is
perhaps most familiar in the context of connected compact orientable surfaces, where it
characterizes the number of handles in the surface: $\chi=2(1-\#)$ for $\#$
handles. In higher dimensions it is more difficult to interpret, but there are
a few basic intuitions.  The Euler characteristic of the hypersphere is $2$ in even dimensions and 0 in odd dimensions. In fact, the Euler
characteristic of an odd-dimensional manifold is always zero. The Euler
characteristic of the union of two disjoint manifolds is the sum of the Euler
characteristics of the individual manifolds, and that of the product of two
manifolds is the product of the Euler characteristics. This means that a
manifold made of many disconnected sphere-like components will have a large
positive Euler characteristic. A manifold with many hyper-handles will have a
large negative Euler characteristic. And no matter the Euler characteristic of
a manifold, the Euler characteristic of its product with the circle $S^1$ is
zero.

The canonical method for computing the Euler characteristic is to construct a
complex on the manifold in question, which is a higher-dimensional
generalization of a polygonal tiling. Then $\chi$ is given by an alternating
sum over the number of cells of increasing dimension, which for 2-manifolds
corresponds to the number of vertices, minus the number of edges, plus the
number of faces. Morse theory offers another way to compute the Euler
characteristic of a manifold $\Omega$ using the statistics of stationary points
in a function $H:\Omega\to\mathbb R$ \cite{Audin_2014_Morse}. For functions $H$
without any symmetries with respect to the manifold, the surfaces of gradient
flow between adjacent stationary points form a complex. The alternating sum
over cells becomes an alternating sum over the count of stationary points of
$H$ with increasing index, or
\begin{equation}
  \chi(\Omega)=\sum_{i=0}^N(-1)^i\mathcal N_H(\text{index}=i)
\end{equation}
Conveniently, we can express this sum as an integral over the manifold
using a small variation on the Kac--Rice formula for counting stationary
points \cite{Kac_1943_On, Rice_1939_The}. Since the sign of the determinant of the Hessian matrix of $H$ at a
stationary point is equal to its index, if we count stationary points including
the sign of the determinant, we arrive at the Euler characteristic, or
\begin{equation} \label{eq:kac-rice}
  \chi(\Omega)=\int_\Omega d\mathbf x\,\delta\big(\nabla H(\mathbf x)\big)\det\operatorname{Hess}H(\mathbf x)
\end{equation}
When the Kac--Rice formula is used to calculate the total number stationary
points, one must take pains to eliminate the sign of the determinant
\cite{Fyodorov_2004_Complexity}. Here it is correct to preserve it.

We need to choose a function $H$ for our calculation. Because $\chi$ is
a topological invariant, any choice will work so long as it does not have degenerate stationary points on the manifold, i.e., that it is a Morse function, and that it does not share some
symmetry with the underlying manifold, i.e., that it satisfies the Smale condition. Because our manifold is random and has no symmetries, we can take a simple height function $H(\mathbf
x)=\mathbf x_0\cdot\mathbf x$ for some $\mathbf x_0\in\mathbb R^N$ with
$\|\mathbf x_0\|^2=N$. We call $H$ a height function because when $\mathbf x_0$ is
interpreted as the polar axis of a spherical coordinate system, $H$ gives the height on the sphere relative to the equator.

We treat the integral over the implicitly defined manifold $\Omega$ using the
method of Lagrange multipliers. We introduce one multiplier $\omega_0$ to
enforce the spherical constraint and $M$ multipliers $\omega_k$ for $k=1,\ldots,M$ to enforce the $M$ constraints, resulting in the Lagrangian
\begin{equation} \label{eq:lagrangian}
  L(\mathbf x,\pmb\omega)
  =H(\mathbf x)+\frac12\omega_0\big(\|\mathbf x\|^2-N\big)
  +\sum_{k=1}^M\omega_k\big(V_k(\mathbf x)-V_0\big)
\end{equation}
The integral over the solution manifold $\Omega$ in \eqref{eq:kac-rice} becomes
\begin{equation} \label{eq:kac-rice.lagrange}
  \chi(\Omega)=\int_{\mathbb R^N} d\mathbf x\int_{\mathbb R^{M+1}}d\pmb\omega
  \,\delta\big(\partial L(\mathbf x,\pmb\omega)\big)
  \det\partial\partial L(\mathbf x,\pmb\omega)
\end{equation}
where $\partial=[\frac\partial{\partial\mathbf x},\frac\partial{\partial\pmb\omega}]$
is the vector of partial derivatives with respect to all $N+M+1$ variables.
This expression is now in a form where standard techniques from the mean-field
theory of disordered systems can be applied to average over the random constraint functions and evaluate the integrals to leading order in large $N$.

\begin{figure}[tb]
  \includegraphics{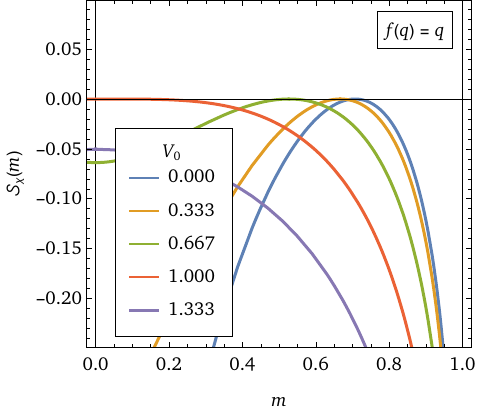}
  \hspace{-3.5em}
  \includegraphics{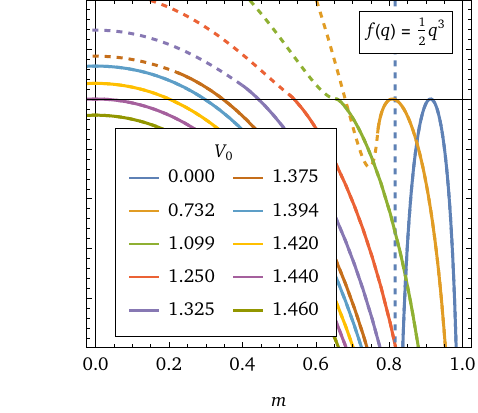}

  \caption{
    \textbf{Effective action for the Euler characteristic.}
    The action \eqref{eq:S.m} as a function of $m=\frac1N\mathbf x\cdot\mathbf
    x_0$ for pure polynomial constraints and a selection of target values
    $V_0$. Dashed lines depict $\operatorname{Re}\mathcal S_\chi$ when its
    imaginary part is nonzero. In both plots $\alpha=\frac12$. \textbf{Left:}
    With linear functions there are two regimes. For small $V_0$, there are
    maxima at $m=\pm m_*$ where the action is zero, while for $V_0>
    V_{\text{\textsc{sat}}\ast}=1$ the action is negative everywhere.
    \textbf{Right:} With nonlinear functions there are other possible regimes.
    For small $V_0$, there are maxima at $m=\pm m_*$ but the real part of the
    action is maximized at $m=0$ where the action is complex. For larger
    $V_0\geq V_\text{on}\simeq1.099$ the maxima at $m=\pm m_*$ disappear. For
    $V_0\geq V_\text{sh}\simeq1.394$ larger still, the action becomes real
    everywhere. Finally, for $V_0>V_\text{\textsc{sat}}\simeq1.440$ the action
    is negative everywhere.
  } \label{fig:action}
\end{figure}

Details of this calculation can be found in Appendix~\ref{sec:euler}. The
result is the reduction of the average Euler characteristic to an integral over
a single order parameter $m=\frac1N\mathbf x\cdot\mathbf x_0$ of the form
\begin{equation}
  \overline{\chi(\Omega)}=\left(\frac{N}{2\pi}\right)^{\frac12}\int dm\,g(m)\,e^{N\mathcal S_\chi(m)}
\end{equation}
where $g(m)$ is a prefactor of order $N^0$ and $\mathcal S_\chi(m)$ is an
effective action defined by
\begin{equation} \label{eq:S.m}
  \mathcal S_\chi(m)
  =-\frac\alpha2\bigg[
    \log\left(
      1-\frac{f(1)}{f'(1)}\frac{1+\frac m{R_m}}{1-m^2}
    \right)
    +\frac{V_0^2}{f(1)}\left(
      1-\frac{f'(1)}{f(1)}\frac{1-m^2}{1+\frac m{R_m}}
    \right)^{-1}
  \bigg]
  +\frac12\log\left(-\frac m{R_m}\right)
\end{equation}
Here we have introduced the ratio $\alpha=M/N$ between the number of equations
and the number of variables, and $R_m$ is a function of $m$ given by
\begin{equation} \label{eq:rs}
  \begin{aligned}
    R_m
    \equiv\frac{-m(1-m^2)}{2[f(1)-(1-m^2)f'(1)]^2}
    \Bigg[
      \alpha V_0^2f'(1)
      +(2-\alpha)f(1)\left(\frac{f(1)}{1-m^2}-f'(1)\right) \quad \\
    \quad+\alpha\sqrt{
      \tfrac{4V_0^2}\alpha f(1)f'(1)\left[\tfrac{f(1)}{1-m^2}-f'(1)\right]
      +\left[\tfrac{f(1)^2}{1-m^2}-\big(V_0^2+f(1)\big)f'(1)\right]^2
    }
  \Bigg]
  \end{aligned}
\end{equation}
The effective action \eqref{eq:S.m} is plotted in Fig.~\ref{fig:action} for a selection of
parameters. To finish evaluating the integral by the saddle-point
approximation, the action should be maximized with respect to $m$. If $m_*$ is
such a maximum, then the resulting average Euler characteristic is
$\overline{\chi(\Omega)}\propto e^{N\mathcal S_\chi(m_*)}$. In the next
subsection we examine the maxima of $\mathcal S_\chi$ and their properties as the
parameters are varied.

\subsection{Features of the effective action}

The order parameter $m$ is the overlap of the
configuration $\mathbf x$ with the height axis $\mathbf x_0$. Therefore, the
value $m$ that maximizes this action can be understood as the latitude on the
sphere at which most of the contribution to the Euler characteristic is made.\footnote{
The order parameter $m$ may resemble the magnetization that appears in
problems that have a signal or spike, where it gives the overlap of a
configuration with the hidden signal. Here $\textbf x_0$ is no signal, but a
direction chosen uniformly at random and with no significance to the set of solutions. Here, if a feature of the
action is present at some value $m$, it should be interpreted as indicating
that, with overwhelming probability, typical configurations contributing to that feature have an overlap $m$ with a typical
point in configuration space. For instance, for $m$
sufficiently close to 1, $\mathcal S_\chi(m)$ is always negative, which is a
result of the absence of any stationary points contributing to the Euler
characteristic at those overlaps. Given a random height axis $\mathbf x_0$, the
nearest point to $\mathbf x_0$ on the solution manifold will be the absolute
maximum of the height function, and therefore will contribute to the Euler
characteristic. Hence the region of negative action in
the vicinity of $m=1$ implies there is a typical minimum distance between the
solution manifold and a randomly drawn point in configuration space, and that
it is vanishingly unlikely to draw a point in configuration space uniformly at
random and find it any closer to the solution manifold than this.
  Other properties of the set of solutions could be studied by drawing $\mathbf
  x_0$ from an alternative distribution, like the Boltzmann distribution of the
  cost function, from the set of its stationary points, or from the solution
  manifold itself. While the value of the Euler characteristic would not
  change, the dependence of the effective action on $m$ would change.
}
The action $\mathcal S_\chi$ is extremized with respect to $m$ at $m=0$ or at $m=\pm m_*$ for
\begin{equation}
  m_*=\sqrt{1-\frac{\alpha}{f'(1)}\big(V_0^2+f(1)\big)}
\end{equation}
At these latter extrema, $\mathcal S_\chi(\pm m_*)=0$. Zero action implies that
$\overline{\chi(\Omega)}$ does not vary exponentially with $N$, and in fact we
show in Appendix~\ref{sec:prefactor} that the contribution from these
extrema is $1+o(N^0)$ at $-m_*$ and $(-1)^{N-M-1}+o(N^0)$ at $+m_*$, so that
their sum is $2$ in even dimensions and $0$ in odd dimensions. When these
extrema exist and maximize the action, this result is consistent with the
topology of an $N-M-1$ sphere.

If this solution were always well-defined, it would
vanish when the argument of the square root vanishes for
\begin{equation}
  V_0^2>V_{\text{\textsc{sat}}\ast}^2\equiv\frac{f'(1)}\alpha-f(1)
\end{equation}
This corresponds precisely to the satisfiability transition found in previous
work by a replica symmetric analysis of the cost function \eqref{eq:cost}
\cite{Fyodorov_2019_A, Fyodorov_2020_Counting, Fyodorov_2022_Optimization,
Tublin_2022_A, Vivo_2024_Random}. However, the action is not clearly defined in
the entire range $m^2<1$: it becomes complex in the region
$m^2<m_\text{min}^2$ where
\begin{equation}
  m_\text{min}^2
  \equiv1-\frac{f(1)^2}{f'(1)}\times
  \frac{V_0^2(1+\sqrt{1-\alpha})^2-\alpha f(1)}{
    4V_0^2f(1)-\alpha[V_0^2+f(1)]^2
  }
\end{equation}
When $m_*^2<m_\text{min}^2$, the solutions at $m=\pm m_*$ are no longer maxima of
the action. This happens when the target value $V_0$ is larger than an onset value $V_\text{on}$ defined by
\begin{equation}
  V_\text{on}^2\equiv\frac{f(1)}\alpha\left(1-\alpha+\sqrt{1-\alpha}\right)
\end{equation}
Comparing this with the satisfiability transition associated with $m_*$ going to zero, one sees
\begin{equation}
  V_\text{on}^2-V_{\text{\textsc{sat}}\ast}^2
  =\frac1\alpha\left(f'(1)-f(1)-f(1)\sqrt{1-\alpha}\right)
\end{equation}
If $f(q)$ is purely linear, then $f'(1)=f(1)$ and
$V_\text{on}^2>V_{\text{\textsc{sat}}\ast}^2$, so the naïve satisfiability
transition happens first. On the other hand, when $f(q)$ contains powers of $q$
strictly greater than 1, then $f'(1)\geq 2f(1)$ and $V_\text{on}^2\leq
V_{\text{\textsc{sat}}\ast}^2$, so the onset happens first. In situations with
mixed constant, linear, and nonlinear terms in $f$, the order of the
transitions depends on the precise form of $f$.

Now we return to the extremum at $m=0$. As for those at $\pm m_*$, the action
evaluated at this solution is sometimes complex-valued and sometimes
real-valued. For $V_0$ less than a shattering value $V_\text{sh}$ defined by
\begin{equation}
  V_\text{sh}^2\equiv\frac{f(1)}\alpha\left(1-\frac{f(1)}{f'(1)}\right)\left(1+\sqrt{1-\alpha}\right)^2
\end{equation}
the maximum at $m=0$ is complex while for $V_0$ greater than this value the action is real. For purely
linear $f(q)$, $V_\text{sh}=0$ and the action at $m=0$ is always real, though
for $V_0^2<V_{\text{\textsc{sat}}\ast}^2$ it is a minimum rather than a
maximum. Finally, there is another satisfiability transition at
$V_0=V_\text{\textsc{sat}}$ corresponding to the vanishing of the effective
action at the $m=0$ solution, with $\mathcal S(0)=0$. For a generic covariance
function $f$ it is not possible to write an explicit formula for
$V_\text{\textsc{sat}}$, and we calculate it through a numeric
root-finding algorithm.\footnote{
As a check of this calculation, the satisfiability threshold calculated here can be compared with that calculated using the zero-temperature limit of an equilibrium treatment of the cost function \eqref{eq:cost} made in Ref.~\cite{Urbani_2023_A} for the case where $f(q)=\frac12q^2$ and $\alpha=\frac14$. The authors estimate $V_\text{\textsc{sat}}\simeq1.871$, whereas this manuscript predicts $V_\text{\textsc{sat}}=1.867229\dots$, a seeming inconsistency. However, the author of Ref.~\cite{Urbani_2023_A} indicated in private correspondence that this difference is explained by inaccuracy in the numeric \textsc{pde} treatment of the \textsc{frsb} equilibrium problem. Therefore, this manuscript is consistent with the previous work, but the agreement is not precise.
}

When $V_0^2<V_\text{sh}^2$, the solution at $m=0$ is difficult to interpret, since
the action takes a complex value. Such a result could arise from the breakdown
of the large-deviation principle behind the calculation of the effective
action, or it could be the result of a negative Euler characteristic.
To address this ambiguity, we compute also the average of the square of the Euler
characteristic, $\overline{\chi(\Omega)^2}$, with details in
Appendix~\ref{sec:rms}. This has the benefit of always being positive, so that
the saddle-point approach to the calculation at large $N$ does not produce
complex values even when $\overline{\chi(\Omega)}$ is negative. Under the restriction that $f(0)=0$,\footnote{
  This restriction is equivalent to having no random constant term in the
  constraint equations. It provides a simplification here because when it is
  present the replica symmetric (\textsc{rs}) description of this problem can
  have $q_0>0$, and $\overline{\chi(\Omega)^2}\neq[\overline{\chi(\Omega)}]^2$
  always.
} we identify three
saddle points that could contribute to the value of
$\overline{\chi(\Omega)^2}$: two at $\pm m_*$ where
$\frac1N\log\overline{\chi(\Omega)^2}=\frac1N\log\overline{\chi(\Omega)}\simeq0$,
and one at $m=0$ where
\begin{equation}
  \frac1N\log\overline{\chi(\Omega)^2}=2\operatorname{Re}\mathcal S_\chi(0)
\end{equation}
which is consistent with
$\overline{\chi(\Omega)^2}\simeq[\overline{\chi(\Omega)}]^2$. We therefore
conclude that when the effective action is complex-valued, the average Euler
characteristic is negative and its magnitude is given by the real part of the
action.

Such a correspondence, which indicates that the `annealed' calculation
presented here is also representative of typical realizations of the
constraints, is not always true. Sometimes the average squared Euler
characteristic has alternative saddle points for which
$\overline{\chi(\Omega)^2}\neq[\overline{\chi(\Omega)}]^2$, which implies that
average properties will not be typical of most realizations. With our
calculation of the average squared Euler characteristic, we can identify
instabilities of the solution described above toward such replica symmetry
breaking (\textsc{rsb}) solutions. The analysis of these instabilities can be found in Appendix~\ref{sec:rsb.instability}.  We do not explore these \textsc{rsb}
solutions here, except in the context of $M=1$ and the spherical spin glasses
in Section~\ref{sec:ssg}. However, in the phase diagrams of Figures \ref{fig:phases}
and \ref{fig:crossover} we shade the region where our calculation indicates that an instability is present.

\subsection{Topological phases and their interpretation}

\begin{figure}
  \includegraphics[width=0.196\textwidth]{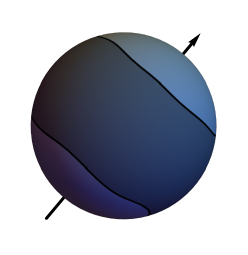}
  \includegraphics[width=0.196\textwidth]{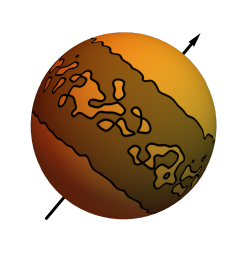}
  \includegraphics[width=0.196\textwidth]{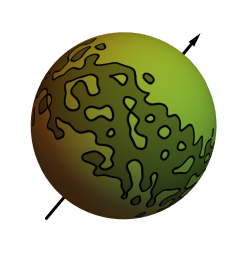}
  \includegraphics[width=0.196\textwidth]{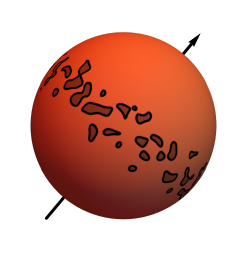}
  \includegraphics[width=0.196\textwidth]{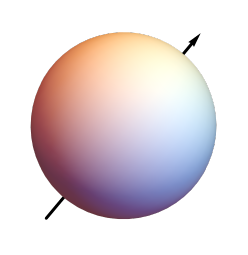}

  \hspace{1.5em}
  \textbf{Regime I}
  \hfill
  \textbf{Regime II}
  \hfill
  \textbf{Regime III}
  \hfill
  \textbf{Regime IV}
  \hfill
  \textbf{Regime V}
  \hspace{1.5em}

  \caption{
    \textbf{Cartoons of the solution manifold in five topological regimes.}
    The solution manifold is shown as a shaded region, and the height axis
    $\mathbf x_0$ is a black arrow. In Regime I, the average Euler
    characteristic is consistent with a manifold with a single simply-connected
    component. In Regime II, holes occupy the equator but the temperate
    regions are topologically simple. In Regime III, holes dominate and the
    edge of the manifold is not necessarily simple. In Regime IV, disconnected
    components dominate. In Regime V, the manifold is empty.
  } \label{fig:cartoons}
\end{figure}

The results of the previous section allow us to unambiguously define distinct
topological phases, which differ depending on the presence or absence of the
local maxima at $m=\pm m_*$, on the presence or absence of the local maximum
at $m=0$, on the real or complex nature of this maximum, and finally on
whether the action is positive or negative. Below we enumerate these regimes,
which are schematically represented in Fig.~\ref{fig:cartoons}.\footnote{
  In the following we characterize regimes by values of
  $\overline{\chi(\Omega)}$. These should be understood as their values in
  \emph{even} dimensions, since in odd dimensions the Euler characteristic is
  always identically zero. We do not expect the qualitative results to change
  depending on the evenness or oddness of the manifold dimension.
} It is not possible to definitively ascertain what structural features of the
solution manifold lead to these average invariants, but we suggest a simplest
interpretation consistent with the calculated properties.

\paragraph{Regime I: \boldmath{$\overline{\chi(\Omega)}=2$}.}

This regime is found when the magnitude of the target value $V_0$ is less than
the onset $V_\text{on}$ and $\operatorname{Re}\mathcal S(0)<0$, so that the
maxima at $m=\pm m_*$ exist and are the dominant contributions to the average
Euler characteristic. Here, $\overline{\chi(\Omega)}=2+o(1)$ for even $N-M-1$,
strongly indicating a topology homeomorphic to the $S^{N-M-1}$ sphere. This regime is the only nontrivial one found with linear covariance $f(q)=q$, where the solution manifold must be a sphere if it is not empty.

\paragraph{Regime II: \boldmath{$\overline{\chi(\Omega)}$} large and negative, isolated contributions at \boldmath{$m=\pm m_*$}.}

This regime is found when the magnitude of the target value $V_0$ is less than the onset $V_\text{on}$, $\operatorname{Re}\mathcal S(0)>0$, and the value of the action at $m=0$ is complex. The dominant contribution to the average Euler characteristic comes from the equator at $m=0$, but the complexity of the action implies that the Euler characteristic is negative.
While the topology of the manifold is not necessarily connected in this
regime, holes are more numerous than components. Since $V_0^2<V_\text{on}^2$,
there are isolated contributions to $\overline{\chi(\Omega)}$ at $m=\pm m_*$.
This implies a temperate band of relative simplicity: given a random point on
the sphere, the nearest parts of the solution manifold are unlikely to have holes or
disconnected components.

\paragraph{Regime III: \boldmath{$\overline{\chi(\Omega)}$} large and negative, no contribution at \boldmath{$m=\pm m_*$}.}

The same as Regime II, but with $V_0^2>V_\text{on}^2$. The solutions at
$m=\pm m_*$ no longer exist, and nontrivial contributions to the Euler
characteristic are made all the way to the edges of the solution manifold.

\paragraph{Regime IV: \boldmath{$\overline{\chi(\Omega)}$} large and positive.}

This regime is found when the magnitude of the target value $V_0$ is greater
than the shattering value $V_\text{sh}$ and $\mathcal S(0)>0$. Above the
shattering transition the effective action is real everywhere, and its value at
the equator is the dominant contribution. Large connected components of the
manifold may or may not exist, but small disconnected components outnumber
holes.\footnote{
  We interpret the large Euler characteristic to indicate a manifold with many (topologically) spherical disconnected components because the manifold is formed by the process of repeatedly taking non-self-intersecting slices of the previous manifold, starting with a sphere. Therefore, an outcome consisting mostly of (topological) spheres seems most plausible. However, a large Euler characteristic is also consistent with a variety of connected product manifolds, among other exotic possibilities. Definitely ruling out such scenarios is not within the scope of this paper.
}

\paragraph{Regime V: \boldmath{$\overline{\chi(\Omega)}$} very small.}

Here $\frac1N\log\overline{\chi(\Omega)}<0$, indicating that the average
Euler characteristic shrinks exponentially with $N$. Under most conditions
we conclude this is the \textsc{unsat} regime where no manifold exists, but
there may be circumstances where part of this regime is characterized by
nonempty solution manifolds that are overwhelmingly likely to have Euler
characteristic zero.

\begin{figure}
  \includegraphics{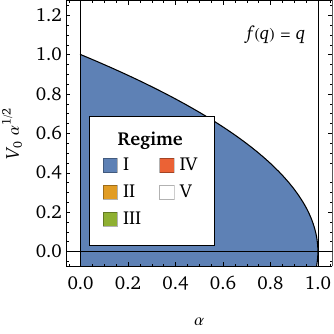}
  \hspace{-3em}
  \includegraphics{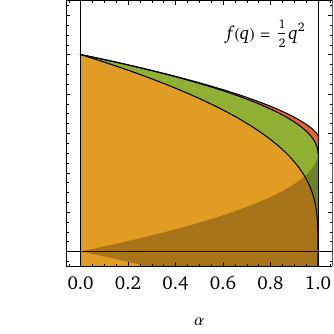}
  \hspace{-3em}
  \includegraphics{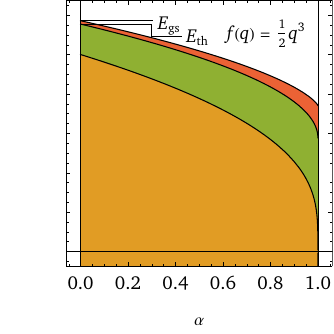}

  \caption{
    \textbf{Topological phase diagram.}
    Topological phases of the problem for three different homogeneous covariance
    functions. The regimes are defined in the text and depicted as cartoons in
    Fig.~\ref{fig:cartoons}. The shaded region in the center panel shows where
    these results are unstable to \textsc{rsb}. In the limit of $\alpha\to0$,
    the behavior of level sets of the spherical spin glasses are recovered: the
    righthand plot shows how the ground state energy
    $E_\text{gs}$ and threshold energy $E_\text{th}$ of the 3-spin spherical model correspond with the limits
    of the satisfiability and shattering transitions in the pure cubic problem. Note that
    for mixed models with inhomogeneous covariance functions, $E_\text{th}$ is
    not the lower limit of $V_\text{sh}$.
  } \label{fig:phases}
\end{figure}

\paragraph{}

The distribution of these phases for situations with homogeneous polynomial
constraint functions is shown in Fig.~\ref{fig:phases}. For purely linear
models, the only two regimes are I and V, separated by a satisfiability
transition at $V_{\text{\textsc{sat}}\ast}$. This is expected: the intersection
of a plane and a sphere is another sphere, and therefore a model of linear
constraints in a spherical configuration space can only produce a solution
manifold consisting of a single sphere, or the empty set. For purely nonlinear
models, regime I does not appear, while the other three nontrivial regimes do.
Regimes II and III are separated by the onset transition at $V_\text{on}$,
while III and IV are separated by the shattering transition at $V_\text{sh}$.
Finally, IV and V are now separated by the satisfiability transition at
$V_\text{\textsc{sat}}$.

An interesting
feature occurs in the limit of $\alpha$ to zero. If $V_0$ is likewise rescaled
in the correct way, the limit of these phase boundaries approaches known
landmark energy values in the pure spherical spin glasses. In particular, the
limit $\alpha\to0$ of the scaled satisfiability transition
$V_\text{\textsc{sat}}\sqrt\alpha$ approaches the ground state energy
$E_\text{gs}$, while the limit $\alpha\to0$ of the scaled shattering
transition $V_\text{sh}\sqrt\alpha$ approaches the threshold energy
$E_\text{th}$. The correspondence between ground state and satisfiability is
expected: when the energy of a level set is greater in magnitude than the
ground state, the level set will usually be empty. The correspondence between
the threshold and shattering energies is also intuitive, since the threshold
energy is typically understood as the point where the landscape fractures into
pieces. However, this second correspondence is only true for the pure spherical
models with homogeneous $f(q)$. For any other model with an inhomogeneous
$f(q)$, $E_\text{sh}^2<E_\text{th}^2$. This may have implications for dynamics
in these mixed models, and we discuss them at length in Section~\ref{sec:ssg}.

\begin{figure}
  \includegraphics{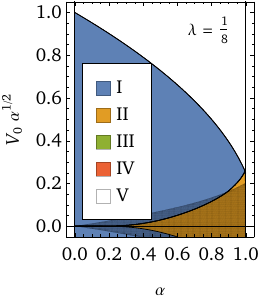}
  \hspace{-2.85em}
  \includegraphics{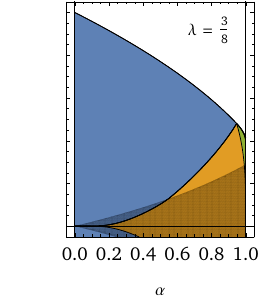}
  \hspace{-2.85em}
  \includegraphics{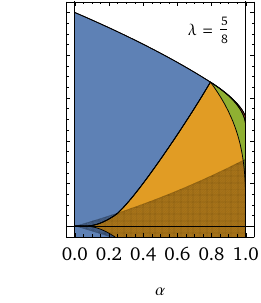}
  \hspace{-2.85em}
  \includegraphics{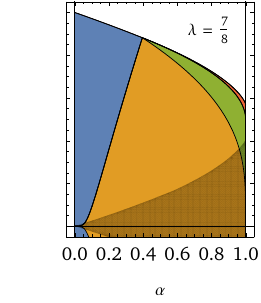}

  \caption{
    \textbf{Linear--quadratic crossover.}
    Topological phases for models with a covariance function
    $f(q)=(1-\lambda)q+\lambda\frac12q^2$ for several values of $\lambda$,
    interpolating between homogeneous linear ($\lambda=0$) and quadratic
    ($\lambda=1$) constraints. The regimes are defined in the text and depicted
    as cartoons in Fig.~\ref{fig:cartoons}. The shaded region on each plot
    shows where these results are unstable to \textsc{rsb}.
  } \label{fig:crossover}
\end{figure}

Rich coexistence between all four regimes occurs in models with mixed linear
and nonlinear constraints. Fig.~\ref{fig:crossover} shows examples of the phase
diagrams for models with a covariance function that interpolates between pure
linear ($\lambda=0$) and pure quadratic ($\lambda=1$). A new phase boundary
appears separating regimes I and II, defined as the point where the real part
of the action at $m=0$ changes from negative to positive. In purely quadratic
case, and in mixed linear and nonlinear cases, there is a substantial region of
the phase diagram shown in Appendix~\ref{sec:rsb.instability} to be susceptible to
\textsc{rsb}, especially for small $V_0$ and large $\alpha$. Future research
into the structure of solutions in this regime is merited.

\section{Implications for the dynamics of spherical spin glasses}
\label{sec:ssg}

When $M=1$ the solution manifold corresponds to the energy
level set of a spherical spin glass with energy density $E=V_0/\sqrt N$. All the
results from the previous sections follow, and can be translated to the spin
glasses by taking the limit $\alpha\to0$ while keeping $E=V_0\alpha^{1/2}$ fixed.\footnote{
  It is plausible that the limit of $N\to\infty$ implicit in the saddle point expansion and the limit of $\alpha\to0$ taken here do not commute, and that $M=1$ should be set from the beginning of the calculation. However, in this case the two procedures do commute. The $\alpha\to0$ limit accomplishes only the elimination of the first term from the effective action \eqref{eq:S.m}, while following Appendix~\ref{sec:euler} with $M=1$ from the outset results in the same term not appearing in the effective action because it is of subleading order in $N$.
} With a little algebra this procedure yields
\begin{align}
  E_\text{on}=\pm\sqrt{2f(1)}
  &&
  E_\text{sh}=\pm\sqrt{4f(1)\left(1-\frac{f(1)}{f'(1)}\right)}
  \label{eq:ssg.energies}
\end{align}
for the onset and shattering energies. The same limit taken for
$V_\text{\textsc{sat}}\alpha^{1/2}$ coincides with the ground state energy
$E_\text{gs}$. In fact, for all energies below the threshold energy
$E_\text{th}$ (where minima become more numerous than saddle points in the
spin glass energy function) the logarithm of the average Euler characteristic
is precisely the complexity of stationary points of the spin glass energy. In
this regime, the Euler characteristic is dominated by contributions coming from
the sphere-like slices of the energy basins directly above minima.

For the pure $p$-spin spherical spin glasses, which have homogeneous covariance functions $f(q)=\frac12q^p$,
the shattering energy is $E_\text{sh}=\sqrt{2(p-1)/p}$, precisely the same as the threshold energy $E_\text{th}$ \cite{Castellani_2005_Spin-glass}. This is intuitive, since threshold energy is widely understood as the
place where level sets are broken into pieces.
However, for general mixed models with inhomogeneous covariance functions the threshold energy is
\begin{equation}
  E_\mathrm{th}=\pm\frac{f(1)[f''(1)-f'(1)]+f'(1)^2}{f'(1)\sqrt{f''(1)}}
\end{equation}
which satisfies $|E_\text{sh}|\leq|E_\text{th}|$. Therefore, as one
descends in energy one will generically meet the shattering energy before
the threshold energy. This is perhaps unexpected, since one might imagine that
where level sets of the energy break into many pieces would coincide with the
largest concentration of shallow minima in the landscape. We see here that this isn't the case.

This fact mirrors another that was made clear recently: when gradient
decent dynamics are run on these models, they will asymptotically reach an
energy above the threshold energy \cite{Folena_2020_Rethinking,
Folena_2021_Gradient, Folena_2023_On}. The old belief that the threshold energy
qualitatively coincides with a kind of shattering of the landscape is one source of the
expectation that the it should coincide with the dynamic limit. Motivated by
our discovery that the actual shattering energy is different from the threshold
energy, we make a comparison of it with existing data on asymptotic dynamics.

\begin{figure}
  \includegraphics{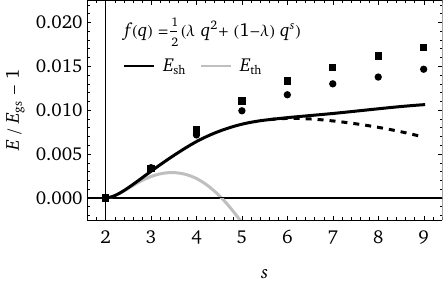}
  \hspace{-0.5em}
  \includegraphics{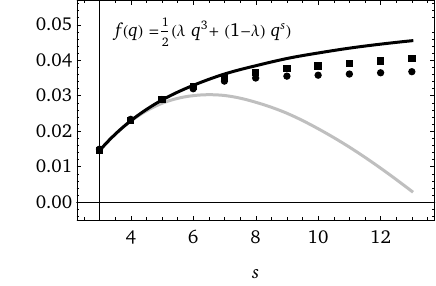}

  \caption{
    \textbf{Is the shattering energy a dynamic threshold?} Comparison of the shattering energy $E_\text{sh}$ with the asymptotic
    performance of gradient descent from a random initial condition in $p+s$
    models with $p=2$ and $p=3$ and varying $s$. The values of $\lambda$ depend on $p$ and $s$ and are taken from \cite{Folena_2023_On}. The points show the asymptotic performance
    extrapolated using two different methods and have unknown uncertainty, from \cite{Folena_2023_On}. Also
    shown is the annealed threshold energy $E_\text{th}$, where marginal minima
    are the most common type of stationary point. The section of $E_\text{sh}$
    that is dashed on the left plot indicates the continuation of the annealed
    result, whereas the solid portion gives the quenched prediction.
  } \label{fig:ssg}
\end{figure}

Measurements of the asymptotic energies reached by dynamics were recently taken in
\cite{Folena_2023_On} for two different classes of models with inhomogeneous
$f(q)$, with
\begin{equation}
  f(q)=\frac12\big[\lambda q^p+(1-\lambda)q^s\big]
\end{equation}
The authors of \cite{Folena_2023_On} studied models with this covariance for $p=2$ and $p=3$ while varying $s$.
In both cases, the relative weight $\lambda$ between the two terms varies with $s$ and was chosen to
maximize a heuristic to increase the chances of seeing nontrivial behavior. The
authors numerically integrated the dynamic mean field theory (\textsc{dmft})
equations for gradient descent in these models from a random initial condition to large but finite time, then attempted to
extrapolate the infinite-time behavior by two different methods. The black
symbols in Fig.~\ref{fig:ssg} show the measurements taken from
\cite{Folena_2023_On}. The difference between the two extrapolations is not
critical here, see the original paper for details. We simply note that the
authors of \cite{Folena_2023_On} did not associate an uncertainty with them,
nor were they confident that they are unbiased estimates of the asymptotic value.

Fig.~\ref{fig:ssg} also shows the shattering and annealed
threshold energies as a function of $s$. The solid lines come from using
\textsl{Mathematica}'s \texttt{Interpolation} function to create a smooth
function $\lambda(s)$ through the values used in \cite{Folena_2023_On}. For the
$2+s$ models for sufficiently large $s$, the ground state is described by a
{\oldstylenums1}\textsc{frsb} order and both the threshold energy and
shattering energy calculated using an annealed average are likely inaccurate
\cite{Auffinger_2022_The}. In Appendix~\ref{sec:1frsb} we calculate the
quenched ground state and shattering energies for these models consistent with
the {\oldstylenums1}\textsc{frsb} equilibrium order. In the left panel of Fig.~\ref{fig:ssg}, the solid line shows the quenched calculation, while the dashed line shows the annealed formula \eqref{eq:ssg.energies}.

Is the shattering energy consistent with the dynamic threshold for gradient
descent from a random initial condition? The evidence in Fig.~\ref{fig:ssg} is
compelling but inconclusive. The difference between the shattering energy and
the extrapolated \textsc{dmft} data is about the same as the difference between
the values predicted by the two extrapolation methods. If both extrapolation
methods suffer from similar systematic biases, it is plausible the true value is the shattering energy. However, better estimates of the
asymptotic values are needed to support or refute this conjecture. This
motivates working to integrate the \textsc{dmft} equations to longer times, or
else look for analytic asymptotic solutions that approach $E_\text{sh}$.

The shattering energy appears consistent with the energy reached by gradient
descent from a uniformly random initial condition, but other algorithms find
minima at other energies. Optimal message passing algorithms were shown to find
configurations at an energy level where another topological property---the
overlap gap property---transitions, and this energy level is believed to bound from below
all polynomial-time algorithms \cite{ElAlaoui_2020_Algorithmic, ElAlaoui_2021_Optimization, Gamarnik_2021-10_The}. On the other hand, physically inspired
modifications of gradient descent---notably, drawing the initial condition from
a nonuniform distribution like the Boltzmann distribution with a finite
temperature---find energy configurations with energies lower than those
found with gradient descent from a uniform initial condition \cite{Folena_2020_Rethinking, Folena_2021_Gradient}. If the
topological transition described in this paper does predict the asymptotic
performance of gradient descent from a uniform initial condition, then it
provides a topological bound from above for the performance of reasonable
algorithms that terminate in minima. It is unknown whether the performance of gradient
descent from better initial conditions, or of other algorithms like simulated
annealing, can be predicted with a topological property.

Finally, a common extension of the spherical spin glasses is to add a
deterministic piece to the energy, sometimes called a signal or a spike. Recent
work argued that gradient descent can avoid being trapped by the minima that typically trap dynamics
and reach the vicinity of the signal if the set of typically trapping minima has
been destabilized by the presence of the signal \cite{SaraoMannelli_2019_Passed,
SaraoMannelli_2019_Who}. The authors of Ref.~\cite{SaraoMannelli_2019_Who}
conjecture based on \textsc{dmft} data for $2+3$ mixed spherical spin glasses that the
typically trapping minima are those at the threshold energy
$E_\text{th}$. However, as discussed above, Ref.~\cite{Folena_2023_On} demonstrated that in signal-free mixed
$p+s$ spherical spin glasses $E_\text{th}$ is not the energy of typical trapping minima, and furthermore that, when $p$ and $s$ are small, the difference between
$E_\text{th}$ and the energy of the actual trapping minima is difficult to resolve with the
current precision of \textsc{dmft} integration schemes. Therefore, it is plausible that the picture described in Ref.~\cite{SaraoMannelli_2019_Who} is correct except that the set of minima that must be destabilised to reach the signal is that at the typical trapping energy of the isotropic problem and not the threshold energy $E_\text{th}$. If the conjecture made in this paper is true, then this typical trapping energy is the shattering energy $E_\text{sh}$.
Comparing the predictions of
Ref.~\cite{SaraoMannelli_2019_Who} to \textsc{dmft} simulations of a model with
better separation between $p$ and $s$ would help resolve this question.

\section{Conclusion}
\label{sec:conclusion}

We have shown how to calculate the average Euler characteristic of the solution
manifold in a simple model of random continuous constraint satisfaction. The results
constrain the topology of this manifold, revealing when it is
connected and trivial, when it is extensive but topologically nontrivial, and
when it is shattered into disconnected pieces.

This calculation has novel implications for the geometry of the energy
landscape in the spherical spin glasses, where it reveals a previously unknown
landmark energy $E_\text{sh}$. This shattering energy is where the topological
calculation implies that the level set of the energy breaks into disconnected
pieces, and differs from the threshold energy $E_\text{th}$ in mixed models.
It's possible that $E_\text{sh}$ is the asymptotic energy reached by gradient descent
from a random initial condition in such models, but the quality of the
currently available data makes this conjecture inconclusive.

Our work also highlights a limitation of using the statistics of
stationary points of an energy or cost function to infer topological properties
of the level sets. In the mixed spherical spin glasses, neither one nor two
stationary point statistics reveal the presence of the topologically significant
energy density $E_\text{sh}$ \cite{BenArous_2019_Geometry,
Folena_2020_Rethinking, Kent-Dobias_2024_Arrangement}. If the shattering energy
is found conclusively to be the dynamic threshold for gradient descent this
failure will be all the more serious. It may be enlightening to return to old
problems of mean-field landscape analysis with this approach in hand, including
in the analysis of the TAP free energy in many spin-glass settings
\cite{Bray_1980_Metastable, Crisanti_1995_Thouless-Anderson-Palmer,
Muller_2006_Marginal}.

This paper has focused on equality constraints, while most existing studies of
constraint satisfaction study inequality constraints \cite{Franz_2016_The,
Franz_2017_Universality, Franz_2019_Critical, Annesi_2023_Star-shaped,
Baldassi_2023_Typical}. To generalize the technique developed in this paper to
such cases is not a trivial extension. The set of solutions to such problems
are manifolds with boundary, and these boundaries are often not smooth. To
study such cases with these techniques will require using extensions of the
Morse theory for manifolds with boundary, and will be the subject of future work.

\paragraph{Acknowledgements}
The authors thank Pierfrancesco Urbani for helpful conversations on these
topics, and Giampaolo Folena for supplying his \textsc{dmft} data for the
spherical spin glasses.

\paragraph{Funding information}
JK-D is supported by FAPESP Young Investigator Grant No.~2024/11114-1. JK-D also received support from the Simons Foundation Targeted Grant to ICTP-SAIFR and a \textsc{DynSysMath} Specific Initiative of the INFN.

\appendix

\section{Details of the calculation of the average Euler characteristic}
\label{sec:euler}

Our starting point is the expression \eqref{eq:kac-rice.lagrange}. To evaluate
the average of $\chi$ over the random constraints, we first translate the
$\delta$-function and determinant to integral form, with
\begin{align}
  \label{eq:delta.exp}
  \delta\big(\partial L(\mathbf x,\pmb\omega)\big)
  &=\int\frac{d\hat{\mathbf x}}{(2\pi)^N}\frac{d\hat{\pmb\omega}}{(2\pi)^{M+1}}
  \,e^{i[\hat{\mathbf x},\hat{\pmb\omega}]\cdot\partial L(\mathbf x,\pmb\omega)}
  \\
  \label{eq:det.exp}
  \det\partial\partial L(\mathbf x,\pmb\omega)
  &=\int d\bar{\pmb\eta}\,d\pmb\eta\,d\bar{\pmb\gamma}\,d\pmb\gamma\,
  e^{-[\bar{\pmb\eta},\bar{\pmb\gamma}]^T\partial\partial L(\mathbf x,\pmb\omega)[\pmb\eta,\pmb\gamma]}
\end{align}
where $\hat{\mathbf x}$ and $\hat{\pmb\omega}$ are ordinary vectors and
$\bar{\pmb\eta}$, $\pmb\eta$, $\bar{\pmb\gamma}$, and $\pmb\gamma$ are
Grassmann vectors. With these expressions substituted into
\eqref{eq:kac-rice.lagrange}, the result is an integral over an exponential
whose argument is linear in the random functions $V_k$.

To make the calculation compact, we introduce
superspace coordinates \cite{DeWitt_1992_Supermanifolds}. An introduction to the use of superspace coordinates in mean field theoretical calculations, including definitions of operators like the superdeterminant using the same conventions as the present article, can be found in Appendix~A of Ref.~\cite{Kent-Dobias_2024_Conditioning}. Introducing the Grassmann indices $\bar\theta_1$
and $\theta_1$, we define the supervectors
\begin{align}
  \pmb\phi(1)=\mathbf x+\bar\theta_1\pmb\eta+\bar{\pmb\eta}\theta_1+\bar\theta_1\theta_1i\hat{\mathbf x}
  &&
  \sigma_k(1)=\omega_k+\bar\theta_1\gamma_k+\bar\gamma_k\theta_1+\bar\theta_1\theta_1i\hat\omega_k
\end{align}
with associated measures
\begin{align}
  d\pmb\phi=d\mathbf x\,\frac{d\hat{\mathbf x}}{(2\pi)^N}\,d\bar{\pmb\eta}\,d\pmb\eta
  &&
  d\pmb\sigma=d\pmb\omega\,\frac{d\hat{\pmb\omega}}{(2\pi)^{M+1}}\,d\bar{\pmb\gamma}\,d\pmb\gamma
\end{align}
The Euler characteristic can be expressed using these supervectors as
\begin{align} \label{eq:kac-rice.super}
  &\chi(\Omega)
  =\int d\pmb\phi\,d\pmb\sigma\,e^{\int d1\,L(\pmb\phi(1),\pmb\sigma(1))} \\
  &=\int d\pmb\phi\,d\pmb\sigma\,\exp\left\{
    \int d1\left[
      H\big(\pmb\phi(1)\big)
      +\frac12\sigma_0(1)\left(\|\pmb\phi(1)\|^2-N\right)
      +\sum_{k=1}^M\sigma_k(1)\Big(V_k\big(\pmb\phi(1)\big)-V_0\Big)
    \right]
  \right\} \notag
\end{align}
where $d1=d\bar\theta_1\,d\theta_1$ is the integration measure over both Grassmann
indices. Since this is an exponential integrand linear in the Gaussian
functions $V_k$, we can take their average to find
\begin{equation} \label{eq:χ.post-average}
  \begin{aligned}
    \overline{\chi(\Omega)}
    =\int d\pmb\phi\,d\pmb\sigma\,\exp\Bigg\{
      \int d1\left[
        H(\pmb\phi(1))
        +\frac12\sigma_0(1)\big(\|\pmb\phi(1)\|^2-N\big)
        -V_0\sum_{k=1}^M\sigma_k(1)
      \right] \\
      +\frac12\int d1\,d2\,\sum_{k=1}^M\sigma_k(1)\sigma_k(2)f\left(\frac{\pmb\phi(1)\cdot\pmb\phi(2)}N\right)
    \Bigg\}
  \end{aligned}
\end{equation}
This is a super-Gaussian integral in the super-Lagrange multipliers $\sigma_k$ with $1\leq k\leq M$.
Performing that integral yields
\begin{equation} \label{eq:pre.hubbard-strat}
  \begin{aligned}
    \overline{\chi(\Omega)}
    &=\int d\pmb\phi\,d\sigma_0\,\exp\Bigg\{
      \int d1\left[
        H(\pmb\phi(1))
        +\frac12\sigma_0(1)\big(\|\pmb\phi(1)\|^2-N\big)
      \right] \\
    &\hspace{5em}-\frac M2V_0^2\int d1\,d2\,f\left(\frac{\pmb\phi(1)\cdot\pmb\phi(2)}N\right)^{-1}
    -\frac M2\log\operatorname{sdet}f\left(\frac{\pmb\phi(1)\cdot\pmb\phi(2)}N\right)
  \Bigg\}
  \end{aligned}
\end{equation}
The supervector $\pmb\phi$ enters this expression as a function only of the
scalar product with itself and with the vector $\mathbf x_0$ inside the
height function $H(\mathbf x)=\mathbf x_0\cdot\mathbf x$. We therefore make a change
of variables to the superoperator $\mathbb Q$ and the supervector $\mathbb M$
defined by
\begin{equation}
  \mathbb Q(1,2)=\frac{\pmb\phi(1)\cdot\pmb\phi(2)}N
  \qquad
  \mathbb M(1)=\frac{\pmb\phi(1)\cdot\mathbf x_0}N
\end{equation}
These new variables can replace $\pmb\phi$ in the integral using a generalized Hubbard--Stratonovich transformation, which yields
\begin{align}
  &\overline{\chi(\Omega)}
    =\frac12\int d\mathbb Q\,d\mathbb M\,d\sigma_0\,
    \left([\operatorname{sdet}(\mathbb Q-\mathbb M\mathbb M^T)]^\frac12+O(N^{-1})\right)
    \,\exp\Bigg\{
    \frac N2\log\operatorname{sdet}(\mathbb Q-\mathbb M\mathbb M^T)
  \label{eq:post.hubbard-strat}
    \\
  &+N\int d1\left[
        \mathbb M(1)
        +\frac12\sigma_0(1)\big(\mathbb Q(1,1)-1\big)
      \right]
    -\frac M2V_0^2\int d1\,d2\,f(\mathbb Q)^{-1}(1,2)
    -\frac M2\log\operatorname{sdet}f(\mathbb Q)
  \Bigg\}
  \notag
\end{align}
where we show the asymptotic value of the prefactor in Appendix~\ref{sec:prefactor}.
To move on from this expression,
we need to expand the superspace notation. We can write
\begin{equation} \label{eq:ops.q}
  \begin{aligned}
    \mathbb Q(1,2)
    &=C-R(\bar\theta_1\theta_1+\bar\theta_2\theta_2)
    -G(\bar\theta_1\theta_2+\bar\theta_2\theta_1)
    -D\bar\theta_1\theta_1\bar\theta_2\theta_2 \\
    &\qquad
    +(\bar\theta_1+\bar\theta_2)H
    +\bar H(\theta_1+\theta_2)
    -(\bar\theta_1\theta_1\bar\theta_2+\bar\theta_2\theta_2\bar\theta_1)i\hat H
    -\bar{\hat H}(\theta_1\bar\theta_2\theta_2+\theta_1\bar\theta_1\theta_1)
  \end{aligned}
\end{equation}
and
\begin{equation}
  \mathbb M(1)
  =m+\bar\theta_1H_0+\bar H_0\theta_1+i\hat m\bar\theta_1\theta_1
  \label{eq:ops.m}
\end{equation}
with associated measures
\begin{align}
  d\mathbb Q
  =dC\,dR\,dG\,\frac{dD}{(2\pi)^2}\,d\bar H\,dH\,d\bar{\hat H}\,d\hat H
  &&
  d\mathbb M=dm\,\frac{d\hat m}{2\pi}\,d\bar H_0\,dH_0
  \label{eq:op.measures}
\end{align}
The order parameters $C$, $R$, $G$, $D$, $m$, and $\hat m$ are ordinary numbers defined by
\begin{align} \label{eq:ops.bos}
  C=\frac{\mathbf x\cdot\mathbf x}N
  &&
  R=-i\frac{\mathbf x\cdot\hat{\mathbf x}}N
  &&
  G=\frac{\bar{\pmb\eta}\cdot\pmb\eta}N
  &&
  D=\frac{\hat{\mathbf x}\cdot\hat{\mathbf x}}N
  &&
  m=\frac{\mathbf x_0\cdot\mathbf x}N
  &&
  \hat m=-i\frac{\mathbf x_0\cdot\hat{\mathbf x}}N
\end{align}
while $\bar H$, $H$, $\bar{\hat H}$, $\hat H$, $\bar H_0$ and $H_0$ are Grassmann numbers defined by
\begin{align} \label{eq:ops.ferm}
  \bar H=\frac{\bar{\pmb\eta}\cdot\mathbf x}N
  &&
  H=\frac{\pmb\eta\cdot\mathbf x}N
  &&
  \bar{\hat H}=-i\frac{\bar{\pmb\eta}\cdot\hat{\mathbf x}}N
  &&
  \hat H=-i\frac{\pmb\eta\cdot\hat{\mathbf x}}N
  &&
  \bar H_0=\frac{\bar{\pmb\eta}\cdot\mathbf x_0}N
  &&
  H_0=\frac{\pmb\eta\cdot\mathbf x_0}N
\end{align}
We can treat the integral over $\sigma_0$ immediately. It gives
\begin{equation} \label{eq:sigma0.integral}
  \int d\sigma_0\,e^{N\int d1\,\frac12\sigma_0(1)(\mathbb Q(1,1)-1)}
  =2\times2\pi\delta(C-1)\,\delta(G+R)\,\bar HH
\end{equation}
This therefore sets $C=1$ and $G=-R$ in the remainder of the integrand, as well
as removing all dependence on $\bar H$ and $H$. With these solutions inserted, the remaining terms in the exponential expand to give
\begin{align} \label{eq:sdet.q}
  &\operatorname{sdet}(\mathbb Q-\mathbb M\mathbb M^T)
  =1+\frac{(1-m^2)D+\hat m^2-2Rm\hat m}{R^2}
  -\frac6{R^4}\bar H_0H_0\bar{\hat H}\hat H
  \\
  &\hspace{5pc}+\frac2{R^3}\left[
    (mR-\hat m)(\bar{\hat H}H_0+\bar H_0\hat H)
    -(D+R^2)\bar H_0H_0
    +(1-m^2)\bar{\hat H}\hat H
  \right] \notag
  \\ \label{eq:sdet.fq}
  &\operatorname{sdet}f(\mathbb Q)
  =1+\frac{Df(1)}{R^2f'(1)}
  +\frac{2f(1)}{R^3f'(1)}\bar{\hat H}\hat H
  \\ \label{eq:inv.q}
  &\int d1\,d2\,f(\mathbb Q)^{-1}(1,2)
  =\frac1{f(1)}\left(1+\frac{R^2f'(1)}{Df(1)}\right)^{-1}
  +2\frac{Rf'(1)}{(Df(1)+R^2f'(1))^2}\bar{\hat H}\hat H
\end{align}
The Grassmann terms in these expressions do not contribute to the effective
action, but will be important in our derivation of the prefactor for the
exponential around the stationary points at $\pm m_*$. The substitution of these expressions into \eqref{eq:post.hubbard-strat} without the Grassmann terms yields
\begin{equation} \label{eq:pre-saddle.characteristic}
  \overline{\chi(\Omega)}
  =\left(\frac N{2\pi}\right)^2\int dR\,dD\,dm\,d\hat m\,g(R,D,m,\hat m)\,e^{N\mathcal S_\chi(R,D,m,\hat m)}
\end{equation}
where $g$ is a prefactor of $o(N^0)$ detailed in the following appendix, $\mathcal S_\chi$ is an effective action defined by
\begin{equation} \label{eq:euler.action}
  \begin{aligned}
    \mathcal S_\chi(R,D,m,\hat m)
    &=-\hat m-\frac\alpha2\left[
      \log\left(1+\frac{f(1)D}{f'(1)R^2}\right)
      +\frac{V_0^2}{f(1)}\left(1+\frac{f'(1)R^2}{f(1)D}\right)^{-1}
    \right] \\
    &\hspace{7em}+\frac12\log\left(
      1+\frac{(1-m^2)D+\hat m^2-2Rm\hat m}{R^2}
    \right)
  \end{aligned}
\end{equation}
and where we have introduced the ratio $\alpha=M/N$.
The integral \eqref{eq:pre-saddle.characteristic} can be evaluated to leading
order in $N$ by a saddle point approximation. To get the formula \eqref{eq:S.m}
in the main text, we first extremize this expression with respect to $R$, $D$,
and $\hat m$, which take the saddle-point values
\begin{align}
  R=R_m
  &&
  D=-\frac{m+R_m}{1-m^2}R_m
  &&
  \hat m=0
\end{align}
where $R_m$ is given by \eqref{eq:rs} from the main text.

\section{Calculation of the prefactor of the average Euler characteristic}
\label{sec:prefactor}

Because of our convention of including the appropriate factors of $2\pi$ in the
superspace measure, super-Gaussian integrals do not produce such factors in our
derivation. Prefactors to our calculation come from three sources: the
introduction of $\delta$-functions to define the order parameters, integrals
over Grassmann order parameters, and from the saddle point approximation to the
large-$N$ integral. In addition, there are important contributions of a sign of the magnetization at the solution that arise from our super-Gaussian integrations.

\subsection{Contribution from the Hubbard--Stratonovich transformation}
\label{sec:prefactor.hs}

First, we examine the factors arising from the definition of order parameters. This begins by introducing to the integral \eqref{eq:pre.hubbard-strat} the factor of one
\begin{equation}
  1=(2\pi)^3\int d\mathbb Q\,d\mathbb M\,
  \delta\big(N\mathbb Q(1,2)-\pmb\phi(1)\cdot\pmb\phi(2)\big)\,
  \delta\big(N\mathbb M(1)-\mathbf x_0\cdot\pmb\phi(1)\big)
\end{equation}
where three factors of $2\pi$ come from the measures as defined in
\eqref{eq:op.measures}. Converting the $\delta$-function into an exponential
integral yields
\begin{equation}
  \begin{aligned}
    1=\frac12\int d\mathbb Q\,d\mathbb M\,d\tilde{\mathbb Q}\,d\tilde{\mathbb M}\,
    \exp\Bigg\{
      \frac12\int d1\,d2\,\tilde{\mathbb Q}(1,2)\big(N\mathbb Q(1,2)-\pmb\phi(1)\cdot\pmb\phi(2)\big) \qquad \\
      +\int d1\,i\tilde{\mathbb M}(1)\big(N\mathbb M(1)-\mathbf x_0\cdot\pmb\phi(1)\big)
    \Bigg\}
  \end{aligned}
\end{equation}
where the supervectors and measures for $\tilde{\mathbb Q}$ and $\tilde{\mathbb M}$ are defined
analogously to those of $\mathbb Q$ and $\mathbb M$. This is now a super-Gaussian integral in $\pmb\phi$, which can be performed to yield
\begin{equation}
  \begin{aligned}
  \int d\pmb\phi\,1=\frac12\int d\mathbb Q\,d\mathbb M\,d\tilde{\mathbb Q}\,d\tilde{\mathbb M}\,
  \exp\Bigg\{
    \frac N2\int d1\,d2\,\tilde{\mathbb Q}(1,2)\mathbb Q(1,2)
    +N\int d1\,i\tilde{\mathbb M}(1)\mathbb M(1) \quad \\
    -\frac N2\log\operatorname{sdet}\tilde{\mathbb Q}
    -\frac N2\int d1\,d2\,\tilde{\mathbb M}(1)\tilde{\mathbb Q}^{-1}(1,2)\tilde{\mathbb M}(2)
  \Bigg\}
  \end{aligned}
\end{equation}
We can perform the remaining super-Gaussian integral in $\tilde{\mathbb M}$ to find
\begin{equation}
  \begin{aligned}
    \int d\pmb\phi\,1=
    \frac12\int d\mathbb Q\,d\mathbb M\,d\tilde{\mathbb Q}\,
    (\operatorname{sdet}\tilde{\mathbb Q}^{-1})^{-\frac12}
    \exp\Bigg\{
      -\frac N2\log\operatorname{sdet}\tilde{\mathbb Q}
      \hspace{5em} \\
      +\frac N2\int d1\,d2\,\tilde{\mathbb Q}(1,2)\big[
        \mathbb Q(1,2)-\mathbb M(1)\mathbb M(2)
      \big]
    \Bigg\}
  \end{aligned}
\end{equation}
The integral over $\tilde{\mathbb Q}$ can be evaluated to leading order using the saddle point
method. The integrand is stationary at $\tilde{\mathbb Q}=(\mathbb Q-\mathbb
M\mathbb M^T)^{-1}$, and substituting this into the above expression results in the term
$\frac12\log\det(\mathbb Q-\mathbb M\mathbb M^T)$ in the effective action from
\eqref{eq:post.hubbard-strat}. The saddle point also yields a prefactor of the form
\begin{equation} \label{eq:supermatrix.saddle}
  \left(\operatorname{sdet}_{\{1,2\},\{3,4\}}\frac{\partial^2\frac12\log\operatorname{sdet}\tilde{\mathbb Q}}{\partial\tilde{\mathbb Q}(1,2)\partial\tilde{\mathbb Q}(3,4)}\right)^{-\frac12}
  =\left(\operatorname{sdet}_{\{1,2\},\{3,4\}}\tilde{\mathbb Q}^{-1}(3,1)\tilde{\mathbb Q}^{-1}(2,4)\right)^{-\frac12}
  =1
\end{equation}
where the final superdeterminant is identically 1 for any superoperator $\tilde{\mathbb Q}$, not just its saddle-point value.\footnote{
  The subscript notation in \eqref{eq:supermatrix.saddle} indicates which superindices of the four-index superoperator associated with the Hessian belong to the domain and codomain, analogous to writing $\det A=\det_{ij}A_{ij}$ for a two-index complex-valued operator. In this case, the domain is indexed by $\{3,4\}$ and the codomain is indexed by $\{1,2\}$.
}
The Hubbard--Stratonovich transformation therefore contributes a factor of
\begin{equation}
  \frac12\operatorname{sdet}(\mathbb Q-\mathbb M\mathbb M^T)^{\frac12}
  =\frac12[(C-m^2)(D+\hat m^2)+(R-m\hat m)^2]^\frac12G^{-1}
\end{equation}
to the prefactor at the largest order in $N$.

\subsection{Sign of the prefactor}

The superspace notation papers over some analytic differences between branches
of the logarithm that are not important for determining the saddle point but
are important to getting correctly the sign of the prefactor. For instance, consider the superdeterminant of $\mathbb Q$ from \eqref{eq:ops.q} (dropping the fermionic order parameters for a moment for brevity),
\begin{equation}
  \operatorname{sdet}\mathbb Q=\frac{CD+R^2}{G^2}
\end{equation}
The numerator and denominator arise from the determinant in the sector of ordinary number and Grassmann number basis elements for the superoperator, respectively. In our calculation, such superdeterminants appear after Gaussian integrals, like
\begin{equation}
  \int d\pmb\phi\,\exp\left\{-\frac12\int d1\,d2\,\pmb\phi(1)\mathbb Q(1,2)\pmb\phi(2)\right\}
  =(\operatorname{sdet}\mathbb Q)^{-\frac12}
  =(CD+R^2)^{-\frac12}G
\end{equation}
Here we emphasize that in the expanded result of the integral, the factor from
the denominator of the superdeterminant enters not as $(G^2)^{\frac12}=|G|$ but as
$G$, including its sign. Therefore, when we write in the effective action
$-\frac12\log\operatorname{sdet}\mathbb Q$, we should really be writing
\begin{equation}
  \int d\pmb\phi\,\exp\left\{-\frac12\int d1\,d2\,\pmb\phi(1)\mathbb Q(1,2)\pmb\phi(2)\right\}
  =\operatorname{sign}(G)e^{-\frac12\log\operatorname{sdet}\mathbb Q}
\end{equation}
In our calculation in Appendix \ref{sec:euler} we elide this several times,
and accumulate $M$ factors of $\operatorname{sign}(-Gf'(C))=\operatorname{sign}(-G)$ from the Gaussian
integral over Lagrange multipliers and $N$ factors of $\operatorname{sign}(-G)$
from the Hubbard--Stratonovich transformation. Since at all saddle points $G=-R$, we have
\begin{equation}
  \operatorname{sign}(R)^{N+M}e^{N\mathcal S_\chi(\tilde{\mathbb Q},\mathbb Q,\mathbb M)}
\end{equation}

\subsection{Contribution from integrating the Grassmann order parameters}

After integrating out the Lagrange multiplier enforcing the spherical
constraint in \eqref{eq:sigma0.integral}, the Grassmann variables $\bar H$ and
$H$ are eliminated from the integrand. This leaves dependence on $\bar{\hat
H}$, $\hat H$, $\bar H_0$, and $H_0$. Expanding the contributions from
\eqref{eq:sdet.q}, \eqref{eq:sdet.fq}, and \eqref{eq:inv.q}, the total contribution to the action is given by
\begin{equation}
  \int d\bar{\hat H}\,d\hat H\,d\bar H_0\,dH_0\,\exp\left\{
    N\begin{bmatrix}
      \bar{\hat H} & \bar H_0
    \end{bmatrix}
    \begin{bmatrix}
      h_1 & h_2 \\
      h_2 & h_3
    \end{bmatrix}
    \begin{bmatrix}
      \hat H \\
      H_0
    \end{bmatrix}
    +Nh_4\bar{\hat H}\hat H\bar H_0H_0
  \right\}
  =N^2(h_1h_3-h_2^2)+Nh_4
\end{equation}
where
\begin{align}
  &h_1=\frac1R\left(
    \frac{1-m^2}{D(1-m^2)+R^2-2Rm\hat m+\hat m^2}
    -\alpha\frac{Df(1)^2+R^2f'(1)[V_0^2+f(1)]}{[Df(1)+R^2f'(1)]^2}
  \right)
  \\
  &h_2=\frac1R\frac{Rm-\hat m}{D(1-m^2)+R^2-2Rm\hat m+\hat m^2}
  \qquad
  h_3=-\frac1R\frac{D+R^2}{D(1-m^2)+R^2-2Rm\hat m+\hat m^2}
  \\
  &h_4=-\frac1{R^2}\frac1{D(1-m^2)+R^2-2Rm\hat m+\hat m^2}
\end{align}
The contribution to the prefactor at leading order in $N$ is therefore
\begin{equation}
  \frac{N^2}{R^2[D(1-m^2)+R^2-2Rm\hat m+\hat m^2]}\left(
    \alpha\frac{(D+R^2)[Df(1)^2+R^2f'(1)[V_0^2+f(1)]]}{
      [Df(1)+R^2f'(1)]^2
    }
    -1
  \right)
\end{equation}

\subsection{Contribution from the saddle point approximation}

We now want to evaluate the prefactor for the asymptotic value of
$\overline{\chi(\Omega)}$. From the previous sections, the definition of the
measures $d\mathbb Q$ and $d\mathbb M$ in \eqref{eq:op.measures}, and the
integral over $\sigma_0$ of \eqref{eq:sigma0.integral}, we can now see that the
function $g(R,D,m,\hat m)$ of \eqref{eq:pre-saddle.characteristic} is given by
\begin{equation}
  \begin{aligned}
    g(R,D,m,\hat m)
    =-
    \frac{\operatorname{sign}(R)^{N+M}}{R^3[D(1-m^2)+R^2-2Rm\hat m+\hat m^2]^{\frac12}}
    \hspace{3pc} \\
    \times\left(
      \alpha\frac{(D+R^2)[Df(1)^2+R^2f'(1)[V_0^2+f(1)]]}{
        [Df(1)+R^2f'(1)]^2
      }
      -1
    \right)
  \end{aligned}
\end{equation}
In regime I, there are two saddle points of the integrand that
contribute to the asymptotic value of the integral, at $m=\pm m_*$ with $R=-m_*$, $D=0$, and $\hat m=0$. At this saddle point $\mathcal S_\chi=0$. We can therefore write
\begin{equation}
  \overline{\chi(\Omega)}
  =\sum_{m=\pm m_*}
    g(-m,0,m,0)\big[\det\partial\partial\mathcal S_\chi(-m,0,m,0)\big]^{-\frac12}
\end{equation}
where here $\partial=[\frac{\partial}{\partial R},\frac{\partial}{\partial D},\frac\partial{\partial m},\frac\partial{\partial \hat m}]$ is the
vector of derivatives with respect to the remaining order parameters. For both of the two saddle points, the determinant of the Hessian of the effective action evaluates to
\begin{equation}
  \det\partial\partial\mathcal S_\chi
  =\left[\frac1{(m_*)^4}\left(1-\frac{\alpha[V_0^2+f(1)]}{f'(1)}\right)\right]^2
\end{equation}
whereas
\begin{equation}
  g(\mp m_*,0,\pm m_*,0)
  =\frac{(\mp1)^{N+M+1}}{(m_*)^4}\left(1-\frac{\alpha[V_0^2+f(1)]}{f'(1)}\right)
\end{equation}
The saddle point at $m=-m_*$, characterized by minima of the height function, always contributes with a positive term. On the other hand, the saddle point with $m=+m_*$, characterized by maxima of the height function, contributes with a sign depending on if $N+M+1$ is even or odd. This follows from the fact that minima, with an index of 0, have a positive contribution to the sum over stationary points, while maxima, with an index of $N-M-1$, have a contribution that depends on the dimension of the manifold.

We have finally that, in regime I,
\begin{equation}
  \overline{\chi(\Omega)}=1+(-1)^{N+M+1}+O(N^{-1})
\end{equation}
When $N+M+1$ is odd, this evaluates to zero. In fact it must be zero to all
orders in $N$, since for odd-dimensional manifolds the Euler characteristic is
always zero. When $N+M+1$ is even, we have $\overline{\chi(\Omega)}=2$ to
leading order in $N$, as specified in the main text.

\section{The average squared Euler characteristic}
\label{sec:rms}

\subsection{Derivation}

Here we calculate $\overline{\chi(\Omega)^2}$, the average of the squared Euler
characteristic. This is accomplished by taking two copies of the integral \eqref{eq:kac-rice.super}, with
\begin{equation}
  \chi(\Omega)^2
  =\int d\pmb\phi_1\,d\pmb\sigma_1\,d\pmb\phi_2\,d\pmb\sigma_2\,
  e^{\int d1\,[L(\pmb\phi_1(1),\pmb\sigma_1(1))+L(\pmb\phi_2(1),\pmb\sigma_2(1))]}
\end{equation}
The same steps as in the derivation of the Euler characteristic follow. The result is the same as \eqref{eq:post.hubbard-strat}, but with the substitutions of the order parameters with matrices of order parameters,
\begin{align}
  \mathbb Q\mapsto\begin{bmatrix}
    \mathbb Q_{11} & \mathbb Q_{12} \\
    \mathbb Q_{21} & \mathbb Q_{22}
  \end{bmatrix}
  &&
  \mathbb M\mapsto\begin{bmatrix}
    \mathbb M_1 \\
    \mathbb M_2
  \end{bmatrix}
\end{align}
where we have defined
\begin{align}
  \mathbb Q_{ij}(1,2)=\frac1N\pmb\phi_i(1)\cdot\pmb\phi_j(2)
  &&
  \mathbb M_i(1)=\frac1N\pmb\phi_i(1)\cdot\mathbf x_0
\end{align}
Expanding the superindices and applying the Dirac $\delta$-functions implied by
the Lagrange multipliers associated with the spherical constraint (which set
$C_{11}=C_{22}=1$ and $G_{11}=-R_{11}$, $G_{22}=-R_{22}$), we arrive at an
expression
\begin{equation}
  \overline{\chi(\Omega)^2}
  \simeq\int dC_{12}\,dR_{11}\,dR_{12}\,dR_{21}\,dR_{22}\,dD_{11}\,dD_{12}\,dD_{22}\,dG_{12}\,dG_{21}\,dm_1\,dm_2\,d\hat m_1\,d\hat m_2\,e^{N\mathcal S_{\chi^2}}
\end{equation}
with another effective action defined by
\begin{equation}
  \mathcal S_{\chi^2}=-\hat m_1-\hat m_2
  -\frac\alpha2\log\frac{\det A_1}{\det A_2}
  -\frac{\alpha V_0^2}2\begin{bmatrix}
    0 & 1 & 0 & 1
  \end{bmatrix}
  A_1^{-1}
  \begin{bmatrix}
    0 \\ 1 \\ 0 \\ 1
  \end{bmatrix}
  +\frac12\log\frac{\det A_3}{\det A_4}
\end{equation}
with the matrices $A_1$, $A_2$, $A_3$, and $A_4$ given by
\begin{align}
  &
  A_1=\begin{bmatrix}
    D_{11}f'(1) & iR_{11}f'(1) & D_{12}f'(C_{12})+\Delta_{12}f''(C_{12}) & i R_{21}f'(C_{12}) \\
    i R_{11}f'(1) & f(1) & i R_{12}f'(C_{12}) & f(C_{12}) \\
    D_{12}f'(C_{12}) + \Delta_{12}f''(C_{12}) & iR_{12}f'(C_{12}) & D_{22} & iR_{22}f'(1) \\
    iR_{21}f'(C_{12}) & f(C_{12}) & iR_{22}f'(1) & f(1)
  \end{bmatrix}
  \\
  &
  A_2=\begin{bmatrix}
    0 & R_{11}f'(1) & 0 & -G_{21}f'(C_{12}) \\
    -R_{11}f'(1) & 0 & G_{12}f'(C_{12}) & 0 \\
    0 & -G_{12}f'(C_{12}) & 0 & R_{22}f'(1) \\
    G_{21}f'(C_{12}) & 0 & -R_{22}f'(1) & 0
  \end{bmatrix}
  \\
  &
  A_3=\begin{bmatrix}
    1-m_1^2 & i(R_{11}-m_1\hat m_1) & C_{12}-m_1m_2 & i(R_{21}-m_1\hat m_2) \\
    i(R_{11}-m_1\hat m_1) & D_{11}+\hat m_1^2 & i(R_{12}-m_2\hat m_1) & D_{12}+\hat m_1\hat m_2 \\
    C_{12}-m_1m_2 & i(R_{12}-m_2\hat m_1) & 1-m_2^2 & i(R_{22}-m_2\hat m_2) \\
    i(R_{21}-m_1\hat m_2) & D_{12}+\hat m_1\hat m_2 & i(R_{22}-m_2\hat m_2) & D_{22}+\hat m_2^2
  \end{bmatrix}
  \\
  &
  A_4=\begin{bmatrix}
    0 & R_{11} & 0 & -G_{21} \\
    -R_{11} & 0 & G_{12} & 0 \\
    0 & -G_{12} & 0 & R_{22} \\
    G_{21} & 0 & -R_{22} & 0
  \end{bmatrix}
\end{align}
and where $\Delta_{12}=G_{12}G_{21}-R_{12}R_{21}$. The effective action must be
extremized over all the order parameters. We look for solutions in two regimes
that are commensurate with the solutions found for the Euler characteristic.
These correspond to $m_1=m_2=0$ and $C_{12}=0$, and $m_1=m_2=\pm m_*$ and
$C_{12}=1$. We restrict ourselves to cases with $f(0)=0$, which correspond to constraint equations without a random constant term. We find such solutions, and in all cases they have
\begin{align}
  G_{12}=G_{21}=R_{12}=R_{21}=D_{12}=\hat m_1=\hat m_2=0
  \\
  D_{ii}=-\frac{m+R_{ii}}{1-m^2}R_{ii}
  \hspace{2em}
  R_{22}=R_{11}^\dagger
  \hspace{2em}
  R_{11}=R_m
\end{align}
where $\dagger$ denotes the complex conjugate and $R_m$ is the saddle point
solution of \eqref{eq:rs}. Upon substituting these solutions into the
expressions above, we find in both cases that
\begin{equation}
  \mathcal S_{\chi^2}=2\operatorname{Re}\mathcal S_\chi
\end{equation}
as referenced in the main text. This corresponds with
$\overline{\chi(\Omega)^2}\simeq[\overline{\chi(\Omega)}]^2$, justifying the
`annealed' approach we have taken in the rest of the paper.

\subsection{Instability to replica symmetry breaking}
\label{sec:rsb.instability}

However, these solutions are not always the correct saddle point for evaluating
the average squared Euler characteristic. When another solution is dominant,
the dissonance between the average square and squared average indicates the
necessity of a quenched calculation to determine the behavior of typical
samples, and also indicates a likely instability to \textsc{rsb}. We can find these points of
instability by examining the Hessian of the action of the average square of the
Euler characteristic at $m=0$. The stability of this matrix is not sufficient
to determine if our solution is stable, since the many $\delta$-functions
employed in our derivation ensure that the resulting saddle point is never at a
true maximum with respect to some combinations of variables. We rather look for
places where the stability of this matrix changes, indicating 
another solution branching from the existing one. However, we must neglect the
branching of trivial solutions, which occur when $R_m$ goes from real- to
complex-valued.

By examination of the results, it appears that nontrivial \textsc{rsb}
instabilities occur along eigenvectors of the Hessian of $\mathcal S_{\chi^2}$
constrained to the subspace spanned by $C_{12}$, $R_{12}$, $R_{21}$, and $D_{12}$. This may
not be surprising, since these are the parameters that represent nontrivial
correlations between the two copies of the system. We can therefore find the \textsc{rsb} instability by looking for nontrivial zeros of
\begin{equation}
  \det\partial\partial\mathcal S_{\chi^2}\equiv\det\frac{\partial^2\mathcal S_{\chi^2}}{\partial[C_{12},R_{12},R_{21},D_{12}]^2}
\end{equation}
evaluated at the $m=0$ solution described above. The resulting expression is
usually quite heinous and we will not reproduce it in its general form in the text, but there is a regime where a dramatic simplification is
possible. The instability always occurs along the direction
$R_{21}=R_{12}^\dagger$, but when $R_m$ is real, $R_{11}=R_{22}$ and the instability occurs along
the direction $R_{21}=R_{12}$. This allows us to examine a simpler action, and
we find the determinant is proportional to two nontrivial factors, with
\begin{equation} \label{eq:stab.det}
  \det\partial\partial\mathcal S_{\chi^2}=-\frac{2B_1B_2}{[r_*f'(1)]^3[(1+r_*)f(1)-r_*f'(1)]^7}
\end{equation}
If we define
$r_*\equiv\lim_{m\to0}R_m/m$, then the factors $B_1$ and $B_2$ are
\begin{align}
  B_1=[(1+r_*)f(1)]^3-3r_*[(1+r_*)f(1)]^2f'(1)
  +\alpha V_0^2\big[2(1+r_*)f'(0)^2+r_*f'(1)f''(0)\big]
  \quad
  \notag \\
  +\alpha r_*f'(0)^2f'(1)-[r_*f'(1)]^3
  -(1+r_*)f(1)\Big(\alpha\big[f'(0)^2+V_0^2f''(0)\big]-3[r_*f'(1)]^2\Big)
\end{align}
\begin{align}
  &B_2=\big[(1+r_*)f(1)-r_*f'(1)\big]^3[f'(1)^2-\alpha f'(0)^2]
  \Big(f'(1)\big[(1+r_*)f(1)-r_*f'(1)\big]-\alpha f'(0)^2\Big)
  \notag \\
  &\qquad-[\alpha V_0^2r_*f'(1)]^2f''(0)\big[(1+r_*)f'(0)^2+r_*f'(1)f''(0)\big]
  \notag \\
  &\qquad-\alpha V_0^2\big[(1+r_*)f(1)-r_*f'(1)\big]^2
  \bigg[
    (1+r_*)f'(0)^2\big[\alpha f'(0)^2-f'(1)^2\big]
    \notag \\
  &\qquad\qquad+
    r_*f'(1)f''(0)\bigg(
      \alpha f'(0)^2\frac{(1+r_*)f(1)-2r_*f'(1)}{(1+r_*)f(1)-r_*f'(1)}
      -(1-r_*)f'(1)^2
    \bigg)
  \bigg]
\end{align}
As $\alpha$ is increased from zero, the first of these factors to go through
zero represents the instability point. These formulas are responsible for
defining the boundaries of the shaded regions in Fig.~\ref{fig:phases} and
Fig.~\ref{fig:crossover}.

Surprisingly, this approach sees no signal of the
replica symmetry breaking (\textsc{rsb}) transition previously found in
\cite{Urbani_2023_A}. The
instability is predicted to occur when
\begin{equation} \label{eq:vrsb}
  V_0^2>V_\text{\textsc{rsb}}^2
  \equiv\frac{[f(1)-f(0)]^2}{\alpha f''(0)}
  -f(0)-\frac{f'(0)}{f''(0)}
\end{equation}
We conjecture that the \textsc{rsb}
instability found in \cite{Urbani_2023_A} is a trait of the cost function
\eqref{eq:cost}, and is not inherent to the structure of the solution manifold.
Perhaps the best evidence for this is to consider the limit of $M=1$, or
$\alpha\to0$ with $E=V_0\sqrt\alpha$ held fixed, where this problem reduces to
the level sets of the spherical spin glasses. The instability \eqref{eq:vrsb}
implies for the pure spherical 2-spin model with $f(q)=\frac12q^2$ that
$E_\textsc{rsb}=\frac12$, though nothing of note is known to occur in the level
sets of 2-spin model at such an energy.

\section{The quenched shattering energy}
\label{sec:1frsb}

Here we share how the quenched shattering energy is calculated under a
{\oldstylenums1}\textsc{frsb} ansatz. To best make contact with prior work on
the spherical spin glasses, we start with \eqref{eq:χ.post-average}. The
formula in a quenched calculation is almost the same as that for the annealed,
but the order parameters $C$, $R$, $D$, and $G$ must be understood as $n\times
n$ matrices rather than scalars. In principle $m$, $\hat m$, $\omega_0$, $\hat\omega_0$, $\omega_1$, and $\hat\omega_1$ should be considered $n$-dimensional vectors, but since in our ansatz replica vectors are constant we can take them to be constant from the start. Expanding the superspace notation, setting $V_0=E\sqrt{N/M}$, and taking $M=1$, we have
\begin{align}
  &\overline{\log\chi(\Omega)}
  =\lim_{n\to0}\frac\partial{\partial n}\int dC\,dR\,dD\,dG\,dm\,d\hat m\,d\omega_0\,d\hat\omega_0\,d\omega_1\,d\hat\omega_1\,
  \exp N\Bigg\{
    n\hat m
    +\frac i2\hat\omega_0\operatorname{Tr}(C-I)
    \notag \\
  &\hspace{2em}-\omega_0\operatorname{Tr}(G+R)
    -in\hat\omega_1E
    +\frac12\log\det\begin{bmatrix}
      C-m^2 & i(R-m\hat m) \\ i(R-m\hat m) & D-\hat m^2
    \end{bmatrix}
    -\frac12\log G^2
    \notag \\
    &\hspace{2em}-\frac12\sum_{ab}^n\left[
      \hat\omega_1^2f(C_{ab})
      +(2i\omega_1\hat\omega_1R_{ab}+\omega_1^2D_{ab})f'(C_{ab})
      +\omega_1^2(G_{ab}^2-R_{ab}^2)f''(C_{ab})
    \right]
  \Bigg\}
\end{align}
We now
make a series of simplifications. Ward identities associated with the BRST
symmetry possessed by the original action \cite{Annibale_2003_The, Annibale_2003_Supersymmetric, Annibale_2004_Coexistence} indicate that
\begin{align}
  \omega_1D=-i\hat\omega_1R
  &&
  G=-R
  &&
  \hat m=0
\end{align}
Moreover, this problem with $m=0$ has a close resemblance to the complexity of
the spherical spin glasses. In both, at the BRST-symmetric saddle point the
matrix $R$ is diagonal with $R=r_dI$ \cite{Kent-Dobias_2023_How}.
To investigate the shattering energy, we can restrict to solutions with $m=0$
and look for the place where such solutions become complex. Inserting these simplifications, we have up to highest order in $N$
\begin{equation}
  \begin{aligned}
    \overline{\log\chi(\Omega)}
    =\lim_{n\to0}\frac\partial{\partial n}\int dC\,dr_d\,d\hat\omega_0\,d\hat\omega_1\,
    \exp N\Bigg\{
      \frac i2\hat\omega_0\operatorname{Tr}(C-I)
      -in\hat\omega_1E
      \qquad\\
      -i\frac12n\omega_1^*\hat\omega_1r_df'(1)
      -\frac12\sum_{ab}^n
        \hat\omega_1^2f(C_{ab})
      +\frac12\log\det
      \left(\frac{-i\hat\omega_1}{\omega_1^*r_d}C+I\right)
    \Bigg\}
  \end{aligned}
\end{equation}
where $\omega_1^*$ is a constant set by satisfying the extremal equations for $D$.
If we redefine $\hat\beta=-i\hat\omega_1$ and $\tilde r_d=\omega_1^*r_d$, we find
\begin{equation}
  \begin{aligned}
    \overline{\log\chi(\Omega)}
    =\lim_{n\to0}\frac\partial{\partial n}\int dC\,d\hat\beta\,d\tilde r_d\,\hat\omega_0\,
    \exp N\Bigg\{
      \frac i2\hat\omega_0\operatorname{Tr}(C-I)
      +n\hat\beta E
      \qquad\\
      +n\frac12\hat\beta\tilde r_df'(1)
      +\frac12\sum_{ab}^n
        \hat\beta^2f(C_{ab})
      +\frac12\log\det
      \left(\frac{\hat\beta}{\tilde r_d}C+I\right)
    \Bigg\}
  \end{aligned}
\end{equation}
which is exactly the effective action for the supersymmetric complexity in the
spherical spin glasses when in the regime where minima dominate
\cite{Kent-Dobias_2023_How}. As the effective action for the Euler characteristic, this expression is always valid. Following the same steps as in
\cite{Kent-Dobias_2023_How}, we can write the continuum version of this action
for arbitrary \textsc{rsb} structure in the matrix $C$ as
\begin{equation} \label{eq:cont.action}
  \frac1N\overline{\log\chi(\Omega)}=\hat\beta E+\frac12\hat\beta\tilde r_df'(1)
  +\frac12\int_0^1dq\,\left[
    \hat\beta^2f''(q)\chi(q)+\frac1{\chi(q)+\tilde r_d\hat\beta^{-1}}
  \right]
\end{equation}
where $\chi(q)=\int_1^qdq'\int_0^{q'}dq''P(q'')$ and $P(q)$ is the distribution of
off-diagonal elements of the matrix $C$ \cite{Crisanti_1992_The, Crisanti_2004_Spherical, Crisanti_2006_Spherical}. This action must be extremized over
the function $\chi$ and the variables $\hat\beta$ and $\tilde r_d$, under the
constraint that $\chi(q)$ is continuous, and that it has $\chi'(1)=-1$ and
$\chi(1)=0$, necessary for $P$ to be a well-defined probability distribution.

Now the specific form of replica symmetry breaking we expect to see is
important. We want to study the mixed $2+s$ models in the regime where they may
have 1-full \textsc{rsb} in equilibrium \cite{Auffinger_2022_The}. For the Euler characteristic like the
complexity, this will correspond to full \textsc{rsb}, in an analogous way to
{\oldstylenums1}\textsc{rsb} equilibria give a \textsc{rs} complexity. Such order is characterized by a piecewise smooth $\chi$ of the form
\begin{equation}
  \chi(q)=\begin{cases}
    \chi_0(q) & q < q_0 \\
    1-q & q \geq q_0
  \end{cases}
\end{equation}
where
\begin{equation}
  \chi_0(q)=\frac1{\hat\beta}[f''(q)^{-1/2}-\tilde r_d]
\end{equation}
is the function implied by extremizing \eqref{eq:cont.action} over $\chi$ ignoring the continuity and other constraints. The
variable $q_0$ must be chosen so that $\chi$ is continuous. The key difference
between \textsc{frsb} and {\oldstylenums1}\textsc{frsb} in this setting is that
in the former case the ground state has $q_0=1$, while in the latter the ground
state has $q_0<1$.

\begin{figure}
  \centering
  \includegraphics{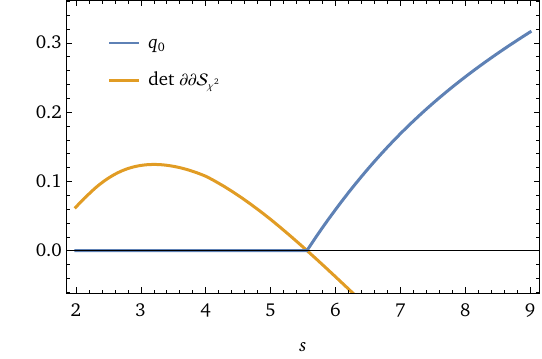}

  \caption{
    \textbf{Self-consistency between \textsc{rsb} instabilities.}
    Comparison between the predicted value $q_0$ for the \textsc{frsb} solution
    at the shattering energy in $2+s$ models and the value of the determinant
    \eqref{eq:stab.det} used in the previous appendix to predict the point of
    \textsc{rsb} instability. The value of $s$ at which $q_0$ becomes nonzero is
    precisely the point where the determinant has a nontrivial zero.
  } \label{fig:rsb}
\end{figure}

We use this action to find the shattering energy in the following way. First,
we know that the ground state energy is the place where the manifold and therefore the average Euler characteristic vanishes. Therefore, setting $\overline{\log\chi(\Omega)}=0$ and solving for $E$ yields a formula
for the ground state energy
\begin{equation}
  E_\text{gs}=-\frac1{\hat\beta}\left\{
    \frac12\hat\beta\tilde r_df'(1)
    +\frac12\int_0^1dq\,\left[
      \hat\beta^2f''(q)\chi(q)+\frac1{\chi(q)+\tilde r_d\hat\beta^{-1}}
    \right]
  \right\}
\end{equation}
This expression can be maximized over $\hat\beta$ and $\tilde r_d$ to find the
correct parameters at the ground state for a particular model. Then, the
shattering energy is found by slowly lowering $q_0$ and solving the combined
extremal and continuity problem for $\hat\beta$, $\tilde r_d$, and $E$ until
$E$ reaches a maximum value and starts to decrease. This maximum is the
shattering energy, since it is the point where the $m=0$ solution becomes complex.
Starting from this point, we take small steps in $s$ and $\lambda_s$,
simultaneously extremizing, ensuring continuity, and maximizing $E$. This draws
out the shattering energy across the entire range of $s$ plotted in
Fig.~\ref{fig:ssg}. The transition to the \textsc{rs} solution occurs when the
value $q_0$ that maximizes $E$ hits zero. We find that the transition between
\textsc{rs} and \textsc{frsb} is precisely predicted by the \textsc{rsb} instability
calculated in Appendix~\ref{sec:rms}, as shown in Fig.~\ref{fig:rsb}.

\bibliography{topology}

\end{document}